          [1999/12/02 1.01 (PWD)]
\documentclass[a4paper,twocolumn]{esapub} 
\usepackage{epsfig}
\usepackage{natbib}

\title{Accretion Disk Spectra of the
Ultra-luminous Compact X-ray Sources in Nearby Spiral
Galaxies and the Super-luminal Jet Sources}
\author{Ken Ebisawa}
\affil{code 662 NASA/GSFC, Greenbelt, MD 20771, USA\footnote{Also
Universities Space Research Association.  {\tt ebisawa@subaru.gsfc.nasa.gov}}}
\author{Aya Kubota,  Tsunefumi Mizuno}
\affil{Department of Physics, The University of Tokyo, 7-3-1 Hongo Bunkyo-ku,
Tokyo 113, Japan}
\author{Piotr \.Zycki}
\affil{N. Copernicus Astronomical Center, Bartycka 18, 00-716 Warsaw, Poland}

\begin{document}


\maketitle

\begin{abstract}
The Ultra-luminous Compact X-ray Sources (ULXs)
in nearby spiral galaxies and the Galactic super-luminal
jet sources share
the common  spectral characteristic  that they have
extremely high disk temperatures which cannot be explained
in the framework of the standard accretion disk model
in the Schwarzschild metric.  We examine several possibilities
to solve this ``too-hot'' disk problem.  In particular, 
we have calculated an extreme Kerr disk model to fit the observed
spectra.  We found that the Kerr disk will become significantly harder
compared to the Schwarzschild disk only when the disk is highly
inclined.
For  super-luminal jet sources,  which are known to be
inclined systems, the Kerr disk model may
work if we choose proper values for the black hole angular momentum.  For the 
ULXs, however, the  Kerr disk interpretation will be problematic,
as it is highly unlikely that their accretion disks are preferentially
inclined.

\end{abstract}

\section{Ultra-luminous Compact X-ray Sources (ULXs)}

Ultra-luminous compact X-ray sources (ULXs) have been found in nearby spiral Galaxies,
with typical 0.5 -- 10 keV luminosities  $10^{39}$ to $10^{40} $ erg s$^{-1}$
(e.g.,  Fabbiano 1988; 
Colbert and Mushotzky 1999; Makishima et al.\ 2000).
These luminosities are too small for AGNs, and  most ULXs are actually located
significantly far from the optical photometric center of the galaxies.  

Significant time variations have been detected from ULXs, and 
their energy spectra are fitted well with optically thick accretion disk models
(Okada et al.\ 1998; Mizuno et al.\ 1999; Makishima et al.\ 2000).
These observational facts  suggest that the ULXs are moderately massive
black holes,  which may be scale-up versions of the Galactic black holes  in the ``high'' state (=soft-state),
in which the soft thermal spectrum is established to be emission from optically thick
accretion disks.
So that the observed luminosity (assuming isotropic emission) 
does not exceed the Eddington limit $L_E = 1.5 \times 10^{38} \,(M/M_\odot)$ erg s$^{-1}$,
the  black hole mass will have to be up to   $\sim 100\; M_\odot$.

In addition to that the energy spectra are fitted with optically thick accretion disk models, 
recently found bimodal spectral transitions  from two ULXs IC342 (Kubota et al.\ 2000)
will further demonstrate their resemblance with the Galactic black hole candidates.

\begin{figure*}
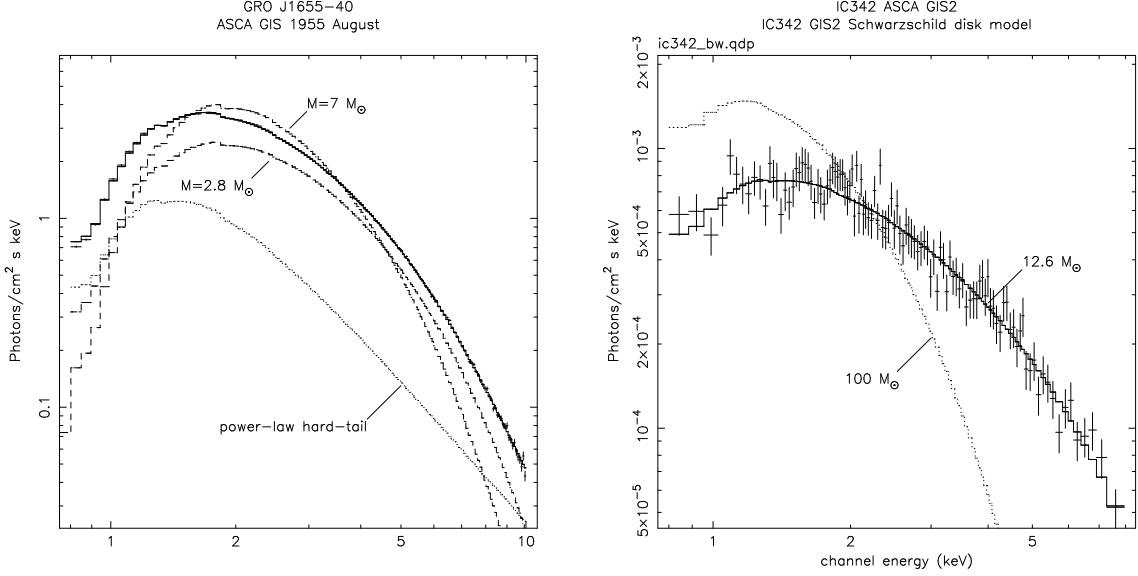

\centerline{
\epsfig{figure=gro1655_fit.ps,angle=270.0,width=7.1cm}\\
\epsfig{figure=ic342_bw.ps,angle=270.0,width=7.1cm}
}
\caption{
ASCA GIS energy spectra of GRO J1655--40
(left) and the Source 1 in IC342 (right panel).
The best-fit Schwarzschild disk models are indicated,
in which the derived masses 
(2.8 $M_\odot$ and 12.6 $M_\odot$
for GRO J1655--40 and IC342 respectively)
are too small compared to
realistic masses.  Schwarzschild disk models with
realistic masses (7 $M_\odot$ and 
100 $M_\odot$ respectively) are also shown, which 
are too soft to explain the observed ``too hot'' accretion disk spectra.
}
\end{figure*}

\section{The  ``Too Hot'' Accretion Disk Problem}

GRS 1915+105 and GRO J1655--40 are the two well-studied  Galactic 
super-luminal jet sources, and the latter is an established
black hole binary with mass measurement of the central source
($M \approx7 M_\odot $; Orosz and Bailyn 1997).
Although these X-ray energy spectra are basically fitted
with an optically thick accretion disk plus power-law,
which is a canonical model for the high-state  black hole
candidates, it has been noticed that the characteristic disk color temperature
of the jet sources can be as high as  $\sim $1.3 -- 2.0 keV (e.g., Belloni et al.\ 1997;
Zhang et al.\ 1997a; Tomsick et al.\ 1999); this  is systematically higher than other non-jet,
well-studied black hole candidates such as Cyg X-1 and   LMC X-3,
whose disk temperatures are  almost always less than $\sim$ 1 keV.

Interestingly, the unusually high disk color temperature, 
$\sim$ 1.5 -- 1.8 keV,  is also observed from the ULXs in  IC342 (Okada et al.\ 1998),  
NGC4565 (Mizuno et al.\ 1999) and other galaxies; in fact this 
seems to be a common spectral characteristic of the ULXs (Makishima et al.\ 2000).

Below, we briefly explain how the characteristic disk temperature is
determined, and why ``the too-high disk temperature'' problem
is of importance.
Let's assume the Schwarzschild metric,
in which case the last stable orbit around the non-rotating black hole
will be  $ 6\; r_{g} (= 6 \;GM/c^2)$.
We consider the accretion disk whose inner disk radius
reaches the last stable orbit, when 
the disk will be hottest and the conversion efficiency is 0.057.
Then the radial dependence
of the ``effective'' temperature of an  optically accretion disk
may be written as,
$$
\sigma T_{eff}^4= \frac{3GM\dot M}{8\pi r^3} R_R(r/r_g), 
$$
where $R_R$ includes minor general relativistic correction
and reduces to $\left(1-\sqrt{6r_g/r}\right)$ in the Newtonian case.
The effective temperature will be maximum at
$r \approx 8 r_g$, where the ``color'' temperature
also reaches the peak, as 
precise disk spectral calculations
indicate that the ratio of the
two temperatures, $T_{col}/T_{eff}$,  is 
virtually constant at $\sim 1.7$ along the disk radius, 
as long as  the $\sim$ 1 -- 10 keV energy band is concerned 
(Shimura and Takahara 1995; Ross and Fabian 1996).

Therefore, the maximum disk color temperature 
in the Schwarzschild metric may be written as
$$
T_{col}^{(max)}  \approx 1.3  {\rm keV} 
\left(\frac{T_{col}/T_{eff}}{1.7}\right)\!\!
\left(\frac{\dot M}{\dot M_{Edd}}\right)^{1/4}\!\!\!\!
\left(\frac{M}{7 \,M_\odot}\right)^{-1/4},
$$
where we define the Eddington mass accretion rate as $\dot M_{Edd} =
3 \times 10^{18} (M/M_\odot)$ g s$^{-1}$,
so that $\dot M / \dot M_{Edd} = 1$ gives the Eddington luminosity.
From this formula, we can see that an optically thick accretion
disk around a 7 $M_\odot$ 
Schwarzschild black hole 
may not be
hotter than the color temperature $\sim $ 1.3 keV.
Note that as the mass
increases the maximum disk temperature decreases with $M^{-1/4}$,
which is the reason that the AGN blue bumps appear in the UV band.

We show examples of the observed ``too-hot'' accretion 
disks in GRO J1655--40 and an ULX in IC342 (Figure 1).
On the left panel of Figure 1, we show an
ASCA GIS spectrum of GRO J1655--40 in 1995 August, when
the source is in one of the brightest states.
The energy spectrum may be fitted with 
an optically thick accretion disk model plus a power-law tail
which extends up to the BATSE energy band with a photon-index of
$\sim  2.5 $ (Zhang et al.\ 1997a).  Contribution of the 
hard-tail in the ASCA band below 10 keV (the yellow line in the panel)
is rather minor and will
not affect the discussion of the optically thick accretion disk model.
We applied a Schwarzschild disk model by Hanawa (1989)
with the distance and inclination angle being fixed at  3.2
 kpc and $70^\circ$ respectively (Orosz and Bailyn 1997), 
and $T_{col}/T_{eff}= 1.7$.
We show the best-fit model which was obtained by
allowing both mass and mass accretion rate to be free;
we have obtained a too small mass $M=2.8 M_\odot$, with 
$\dot M = 2.1 \times 10^{18}$ g s$^{-1}$
($\dot M/\dot M_{Edd}$=0.25).  If we fix the 
mass at  $M=7 M_\odot$, we may  not achieve an acceptable fit, as the
model spectrum is too soft (also shown in  the  panel).

The right panel in Figure 1 indicates an ASCA spectrum
of the ``Source 1'' in IC342 observed in September 1993 when the 
source is in the high state (Okada et al.\ 1998).
The observed luminosity was  $1.1 \times 10^{40}$ erg s$^{-1}$, 
assuming the most likely distance of 4 Mpc (Okada et al.\ 1998 and
references therein) and isotropic emission, 
thus $M > 80 M_\odot$ is expected so that the luminosity
does not exceed the Eddington limit.
We fit the observed spectrum with the 
Schwarzschild disk model by Hanawa (1989), allowing
the mass and mass accretion rate to be free parameters,
assuming  the face-on geometry.
The best-fit model
gives $M=12.6 M_\odot$ and $\dot M = 3 \times 10^{20}$ g s$^{-1}$ 
($\dot M/\dot M_{Edd} = 7.9$).  Obviously, such a super-Eddington
luminosity is very unlikely.  
Even if we assume a  possible minimum distance of 1.5 Mpc (Okada et al.\ 1998),
$\dot M/\dot M_{Edd} \approx 3$ and the super-Eddington problem is not
solved.  Note that changing the inclination angle does not help,
as both $\dot M$and $\dot M_{Edd}$ are  proportional to $\cos i$ and
their ratio is   invariant.  
For a comparison, we have shown
a Schwarzschild disk model spectrum with $M=100 M_\odot$
at the Eddington limit;
it is obvious  that the observed spectrum is much harder than
the model spectrum.

\section{Several Possibilities to Explain the too-hot Disks}


\subsection{Large $T_{col}/T_{eff}$?}

A naive solution to explain the apparently hard
disk spectrum may be to allow $T_{col}/T_{eff}$ to be much
greater than  1.7,
in which case the effective temperature of  the disk remains the
same, but 
significant Comptonization in the upper-layer hardens the disk spectra.
This is an idea by Borozdin et al.\ (1998) who 
adopted $T_{col}/T_{eff} = 2.7$ 
for GRO J1655--40 a posteriori so that  the mass derived from
the X-ray model fitting agrees with the dynamical mass,
as $2.8 M_\odot \times (2.7/1.7)^2 \sim 7 M_\odot$\footnote{Since 
the effective emission radius is proportional to  $M$, 
the disk luminosity $L$ is proportional to $M^2 \, T_{eff}\,^4 =
M^2 \,(T_{eff}/T_{col})\,^4 \,T_{col}\,^4 \equiv A \,T_{col}\,^4$,
in which from X-ray spectral fitting  we can determine  $A$ and $T_{col}$.
Therefore, $M \propto  (T_{col}/T_{eff})^2$.}.

However,  it is  unlikely that $T_{col}/T_{eff} \gg 1.7 $ holds 
universally.  First, the spectral hardening
factor can be calculated by solving the vertical radiative transfer 
in the disk, and several independent calculations (e.g., 
Shimura and Takahara 1995; Ross and Fabian 1996) 
indicate that $T_{col}/T_{eff}  $ is in the range of 1.5 -- 1.9
for reasonable values of the $\alpha$ parameter.  Second, assuming 
$T_{col}/T_{eff} \approx 1.7 $, the black hole mass estimated from the
X-ray spectral fitting agrees fairly well with the 
dynamical mass derived from optical observations
for most high-state black hole  candidates
such as LMC X-3 (Ebisawa et al.\ 1993) and Cyg X-1 (Dotani et al.\ 1997).
If we apply  $T_{col}/T_{eff} = 2.7$ 
for other well-known black hole candidates, we will end up with 
unusually high black hole masses (Shrader and Titarchuk 1999).
Also, application to the accretion disks in neutron star binaries will give
too large masses which are not allowed for neutron stars.


\subsection{Non-standard Disk Structure?}

Instead of the standard optically thick disk
in which all the released gravitational energy is thermalized
and goes into the radiation, if energy transfer
due to advection is introduced, we will have a different 
accretion disk structure and spectrum. Watarai et al.\ (2000, 2001) 
considered the optically thick Advection Dominated
Accretion Flow (ADAF)  model (slim-disk) which may be valid 
near the Eddington limit.  They pointed out that this model
may explain the hard disk spectrum of the super-luminal jet sources
as well as  apparently small inner disk radii. 

According to Watarai et al.\ (2000, 2001), 
$T(r) \propto r^{-0.5}$ holds in the optically thick
ADAF model,  as opposed to  
$T(r) \propto r^{-0.75}$ which is the case for the standard disk.
To examine this hypothesis, we have calculated an accretion disk model with
$T(r) \propto r^{-0.5}$ and applied to the ASCA spectrum
of GRO J1655-40.  As a result, we obtained a very poor fit
with a reduced $\chi^2$ of $\sim$ 40.
This is unfavorable to the optically thick ADAF model, though
more precise spectral calculation and detailed comparison
with the observation will be of interest.

\begin{figure*}
\centerline{\epsfig{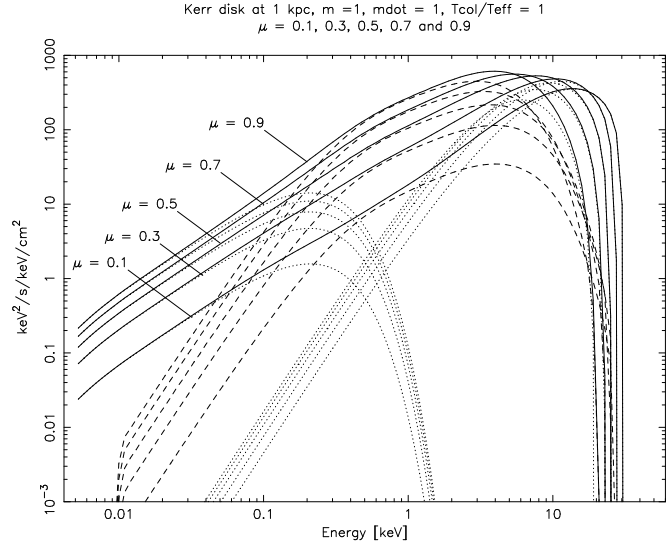}}
\caption{The Kerr accretion 
disk model with an extreme angular momentum (\,$a=0.998$) for the 
inclination angles $\mu \equiv \cos i = 0.1, 0.3, 0.5, 0.7$ and $0.9$.
The distance and  mass
are assumed to be 1 kpc and  1 $M_\odot$ respectively.
The  Eddington luminosity is assumed, and $T_{col}/T_{eff}$ = 1.
Solid lines indicate the total disk spectra, and contributions
from inner (\,$1.26\, r_g < r < 7 \, r_g$), 
middle (\,$7 \, r_g < r < 400 \, r_g$),  and outer parts 
(\,$400 \,r_g < r$) are also plotted seperately
either by dotted  or broken lines.}
\end{figure*}

\subsection{Kerr Metric?}

In the Schwarzschild metric, the last stable orbit
around the black hole is $ 6 \, r_g$, while in the
Kerr metric it can go down to $r_{in} = 1.24\; r_g$
for the prograde case with an extreme angular momentum of $a=0.998$.
If the inner edge of the accretion disk approaches the black hole
accordingly, the local disk temperature can get higher, hence
may explain the too-hot accretion disk spectra 
(Zhang et al.\ 1997b; Makishima et al.\ 2000).
To examine this hypothesis, we have to calculate
the Kerr disk model taken into account full relativisitic 
effects.

Precise Kerr disk model calculations have been carried out by
several authors including  Asaoka (1989), Laor et al.\ (1990)
and Gieli\'nski et al.\ (1999, 2000).  In this paper, we
calculate the Kerr disk spectrum using the transfer function computed 
by Laor et al.\  (1990) for $a=0.998$
(when the conversion efficiency becomes  0.366 and
$\dot M_{Edd} = 4.6 \times 10^{17} (M/M_\odot)$ g s$^{-1}$).
To make  the relativistic effects more apparent, we assume a constant 
$T_{col}/T_{eff}$ and do not take into account the inclination dependent
limb-darkening, which Laor et al.\  (1990) and 
Gieli\'nski et al.\ (1999, 2000) considered.

The calculated energy spectra are shown in Figure 2 for
several inclination angles.  We see that the energy
spectra are strongly dependent on the inclination angle
 in higher  energy bands.
At lower energies, where most of the emission is from the 
outer parts of the disk, the flux decreases with  $\cos i$.
  On the other hand,
the hard emission from 
innermost parts of the disk ($1.26\, r_g < r < 7 \, r_g$)
is {\em enhanced}\/ as the disk is more inclined, due to strong
light bending and Doppler boosting.
When the disk is close to face-on, the contribution from the inner
most part is small, as gravitational red-shift is dominant;
as a result, the total disk spectrum is not so much different
from the Schwarzschild one.
As the disk approaches to edge-on, however, contribution
from the innermost part gets dominant in the highest
band, as a result total disk spectrum becomes much
harder than the Schwarzschild disk.

We have applied the Kerr disk model to
GRO J1655--40 and the Source 1 in IC342.
For GRO J1655--40,  the inclination angle is fixed to  
$i=70^{\circ}$  and  $T_{col}/T_{eff} = 1.7$ is assumed.
If we allow the mass and mass accretion rate to be free, we 
obtain $M = 16.4 M_\odot$ and  $\dot M =  3.5 \times 10^{17}$
g s$^{-1}$ ($\dot M/\dot M_{Edd}=0.046$).
Compared to the fit with Schwarzschild disk model ($M=2.8 M_\odot$),
significant increase of the mass does indicate the spectral
hardening of the Kerr disk model with a large  inclination angle.
In fact,  the derived mass is too large,  
and further tuning of the angular momentum $a$ 
will be required to achieve the realistic mass ($7\, M\odot$).
Gieli\'nski et al.\ (1999, 2000) calculated the Kerr disk model for
several different $a$ values, and found that $a = 0.75$ gives $6.4 M_\odot$
for GRO J1655--40.

The IC 342 fit with  the face-on Kerr disk model
gives (assuming $d$= 4 Mpc)
$M = 27.6 M_\odot$ and $ \dot M =  2.2 \times 10^{20}$
g s$^{-1}$ ($L = 17 \; L_{Edd}$).
Only a factor of $\sim$ 2 increase of the mass is  due to
slight hardening of the face-on Kerr disk model compared to the
Schwarzschild one.  Note that the super-Eddington problem 
is not solved.
If we assume  $i=80^\circ$, we obtain 
$M = 340 M_\odot$ and $ \dot M =  1.7 \times 10^{20}$
g s$^{-1}$ ($L = 1.1 \; L_{Edd}$).
Then the super-Eddington problem will be almost solved, but 
such a large mass seems to be  unlikely.  Also, it is very
unlikely that most of the accretion disks in ULXs are inclined
when seen from the earth.  Therefore, a Kerr disk interpretation of the
ULXs will be problematic, unless considering some other mechanisms to make
the disk spectra apparently harder.

\end{document}